\documentclass[trackchanges]{aastex7}

\usepackage{epstopdf}
  \DeclareGraphicsExtensions{.png,.pdf}

\usepackage{mhchem}
\usepackage{siunitx}
\usepackage{xcolor}
\usepackage[normalem]{ulem}
\pdfoutput=1

\begin{document}

\title{Climates of temperate rocky planets with He-dominated atmospheres}

\author[orcid=0000-0001-9659-5499]{Viviane Kuss}
\affiliation{Department of Earth and Planetary Sciences, Harvard University, 20 Oxford St., Cambridge, MA 02138, USA}
\affiliation{ETH Zurich, Department of Physics, Otto-Stern-Weg 1, 8093, Zurich, Switzerland} 
\email{}

\author[orcid=0000-0003-1127-8334]{Robin Wordsworth} 
\affiliation{Department of Earth and Planetary Sciences, Harvard University, 20 Oxford St., Cambridge, MA 02138, USA}
\affiliation{School of Engineering and Applied Sciences, Harvard University, 20 Oxford St., Cambridge, MA 02138, USA}
\email{}

\author[orcid=0000-0002-8466-5469]{Collin Cherubim}
\affiliation{Department of Earth and Planetary Sciences, Harvard University, 20 Oxford St., Cambridge, MA 02138, USA}
\affiliation{Center for Astrophysics \textbar \ Harvard \& Smithsonian, 60 Garden St., Cambridge, MA 02138, USA}
\email{}

\author[orcid=0009-0003-5970-570X]{Jessica Cmiel}
\affiliation{Department of Earth and Planetary Sciences, Harvard University, 20 Oxford St., Cambridge, MA 02138, USA}
\affiliation{School of Engineering and Applied Sciences, Harvard University, 20 Oxford St., Cambridge, MA 02138, USA}
\email{}

\begin{abstract}

We present radiative-convective modeling of rocky exoplanets with  He-dominated atmospheres and low envelope mass fractions. Helium has a steeper adiabatic temperature profile than \ce{N2} and \ce{H2} as it has fewer degrees of freedom.  Line broadening differences are small, collision-induced absorption (CIA) is not important in \ce{He}-dominated atmospheres, and Rayleigh scattering by He is weak. The combined impact of these effects is that He-dominated atmospheres provide more warming than \ce{N2}-dominated atmospheres but less than \ce{H2}-dominated atmospheres, for the same surface pressure.
For surface pressures in the range 1-20~bar, the habitable zone for rocky planets around M dwarfs with He-dominated atmospheres is narrower than for \ce{H2}-dominated atmospheres. Nonetheless, due to their large scale height compared to \ce{N2} or \ce{CO2}-dominated atmospheres, temperate He-dominated atmospheres are favorable targets for characterization via transit spectroscopy with JWST. 
\end{abstract}


\section{Introduction} \label{sec:introduction}
Sub-Neptunes are a common class of planets, with more than half of Sun-like stars hosting one or more of them \citep{bean2021nature}, although they are notably absent from our solar system.  \citet{hu2015helium} first proposed the existence of sub-Neptunes with He-dominated atmospheres as a result of differentiation during atmospheric escape in He/H$_2$-dominated envelopes. The resulting He-dominated atmospheres may be subject to further mass-loss processes, ultimately leading to planets that lose their primordial envelopes entirely. 
\citet{zhang2023detection} reported the detection of He in the outflows of sub-Neptunes orbiting K dwarfs. Furthermore, a broad population of sub-Neptunes with He-dominated atmospheres at the upper edge of the radius valley was predicted by \citet{cherubim2024strong}, \citet{malsky2023helium}, \citet{gu2023deuterium} and \citet{malsky2020coupled}. Coupled atmosphere-interior simulations predict many of these atmospheres will be oxidized relative to solar composition due to preferential H escape \citep{cherubim2025oxidation}. 

Here, we present the results of the first radiative-convective climate modeling of He-dominated atmospheres. We study the physical effects that cause He to differ from other background gases and explore the possibility of observable He-dominated atmospheres of rocky exoplanets that originated as sub-Neptunes \citep{pierrehumbert2011hydrogen}. 
We give a short overview of our methods and report our key results in Section~\ref{sec:results}. We discuss the implications regarding detectability and atmospheric evolution in Section \ref{sec:discussion}, and explore the possibility of He-dominated ocean worlds. Conclusions are given in Section~\ref{sec:conclusion}.

\section{Method} \label{sec:method}
We use the Planetary Climate Model Line-By-Line (PCM--LBL) one-dimensional radiative-convective model \citep{wordsworth2017transient,wordsworth2021coupled,cmiel2025characterizing} to model He-dominated atmospheres. PCM--LBL is an open-source code that solves the equations of radiative transfer in a plane-parallel atmosphere using line absorption data from the HITRAN database. The model iteratively evolves the temperature towards radiative-convective equilibrium by balancing absorbed shortwave radiation (ASR) and outgoing longwave radiation (OLR). At every timestep, when local super-adiabatic profiles, convective adjustment restores a moist adiabatic profile. Regions of the atmosphere that are stable to convection evolve to radiative equilibrium. Turbulent planetary boundary layer behavior is not taken into account separately from moist convection. Further details are given in \citet{wordsworth2017transient} and \citet{cmiel2025characterizing}. \newline
To adapt PCM--LBL to He-dominated atmospheres, we incorporated new refractive indices for Rayleigh scattering as well as He line broadening and He-\ce{CO2} collision-induced absorption (CIA) data \citep{pierrehumbert2010principles, tan2022h2}. We performed simulations using both a G dwarf spectrum (solar) and an M dwarf spectrum \cite[surface temperature $T_s$ = 3500 K, log $\: g$ = 5.0, and $m$ = 0.0 dex, provided in the Phoenix Library;][]{husser2013new}. The M dwarf spectrum resembles that of LHS 1140 and K2-18, though the temperature is chosen to be closer to K2-18 to maintain consistency across this work, including the transmission-spectrum modeling performed using \verb|PICASO|. This choice is not critical at fixed received stellar instellation flux, as the resulting atmosphere profiles vary only slightly across stellar temperatures $T_s$ from 3100 K to 3500 K and for variations in log $\: g$ between 0.5 and 5.0; in all cases, $T_{surf}$ variations were less than 0.5 K. To elucidate the radiative-convective properties of He atmospheres, we first modeled a hypothetical 1~bar He-dominated atmosphere and compared it to H$_2$- and N$_2$-dominated atmospheres, respectively. In all three cases, our modeled atmospheres contain 1\% \ce{CO2}, condensable \ce{H2O} with fixed tropospheric relative humidity of 0.80 and constant stratospheric molar concentration, and a received stellar instellation flux of 700~\si{W.m^{-2}}. \newline
We next ran the model while varying orbital distance, for the M dwarf case only. 
On Earth-like planets with surface liquid water, the \ce{CO2} concentration may change with orbital distance due to the carbonate-silicate cycle or related feedbacks \citep{walker1981negative}. To account for this possibility, we produced results for three different \ce{CO2} molar concentrations (1\%, 50\%, 95\%). We also ran simulations for two different surface pressures ($p_s$ = 1~bar and $p_s$ = 20~bar). We validated the model results at high temperature and \ce{CO2} molar concentrations by comparing with \citet{wordsworth2013water}.

To explore the detectability of potentially habitable He-dominated atmospheres, we used the open-source code \verb|PICASO| \citep{batalha2019exoplanet}, which simulates ideal transmission spectra for exoplanets.  
We used the planet LHS 1140 b as a reference point, implementing its planetary and stellar properties as described in \citet{cadieux2024new}. LHS 1140 b is predicted by \citet{cherubim2025oxidation} to have a 59\% probability of having at least 30\% atmospheric He-concentration, provided the planet formed in-situ as a sub-Neptune.
Measurements of escaping helium from the planet are consistent with a He-dominated atmosphere (Cherubim et al., in review).
It therefore serves as a useful test case for the exploration of possible He-dominated transit spectra. 
We implemented molar concentrations of 99$\%$ \ce{H2}, \ce{N2}, and He, containing 1$\%$ \ce{CO2} each, similar to the initial atmosphere profiles characterizing the properties of \ce{He} atmospheres. In all cases, variable H$_2$O was included and the surface pressure was 20~bar. He-rich atmospheres with low \ce{H} are expected to be depleted in \ce{CH4} and rich in \ce{CO2}, as described in \citet{hu2015helium} and \citet{cherubim2025oxidation}. \newline
Lastly, we simulated JWST data utilizing the open-source code \verb|PandExo| (\citet{batalha2017pandexo}) with the transmission spectra created by \verb|PICASO| as input. \verb|PandExo| generates the in-transit and out-of-transit models and subsequently utilizes \verb|Pandeia| to apply the corresponding point spread function and account for background noise, ramp noise, read noise, as well as flat-field errors. In this case, we chose NIRSpec PRISM since it covers the desired wavelength range (including our target \ce{CO2} and \ce{H2O} bands).

\section{Results} \label{sec:results}

\begin{figure*}[ht!]
\centering
\includegraphics[scale = 0.37]{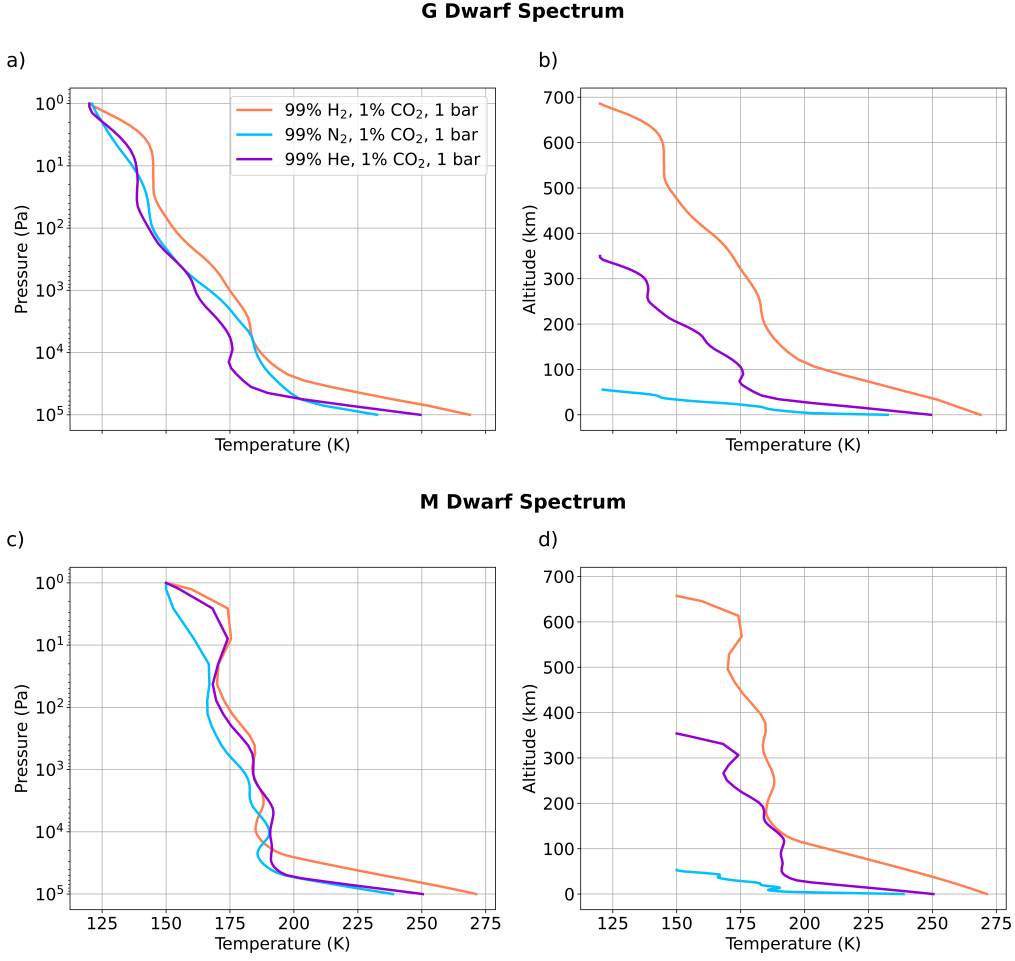}
\caption{Temperature profiles of atmospheres dominated by H$_2$, N$_2$, and He with 1\% CO$_2$ and $p_s$ = 1~bar with a stellar flux of 700~\si{W.m^{-2}}. Figure \textbf{a)} shows temperature vs. pressure while figure \textbf{b)} shows temperature vs. altitude, both for the solar spectrum. Figures \textbf{c)} and \textbf{d)} show the equivalent simulations assuming an M dwarf spectrum with $T_S$ = 3500~K.}
\label{fig:1}
\end{figure*}

The results for the 1~bar modeling given \ce{He}, \ce{H2} and \ce{N2} dominated atmospheres are shown in Figure \ref{fig:1}. For these atmospheres, an incoming stellar flux of 700~\si{W.m^{-2}} is assumed, as well as 1$\%$ \ce{CO2} molar concentration and condensable \ce{H2O} with a relative humidity of 0.80. The altitude-temperature plot illustrates the large differences in scale height for the three atmospheres due to their differing mean molecular masses. The \ce{H2}-\ce{H2} CIA effects are significant and lead to notably higher surface temperature in the \ce{H2}-dominated cases. In contrast, \ce{N2}-\ce{N2} and \ce{He}-\ce{CO2} CIA have an almost negligible effect compared to water vapor at these temperatures. Line broadening differences among \ce{N2}, \ce{He} and \ce{H2} are also small. %Since we implemented an M dwarf stellar spectrum, 
In the G dwarf case, Rayleigh scattering causes a lower surface temperature in the \ce{N2} case (see top left plot in Figure \ref{fig:1}), while the  lower polarizability of \ce{H2} and He leads to weak Rayleigh scattering and higher surface temperature. The effects of Rayleigh scattering are muted for all compositions in the M dwarf case due to the red-shifted spectrum \citep{kasting1993habitable,wordsworth2010gliese}. 

The most important difference between He and \ce{N2} in these simulations comes from the adiabat slope in the convective lower region of the atmosphere (i.e., the troposphere).  
The adiabat slope is determined by $R/c_p$, where $R$ is the specific gas constant and $c_p$ is the specific heat capacity at constant pressure. It can be expressed as $R/c_p = \gamma/(1-\gamma)$, where $\gamma$ is the adiabatic index. In the classical equipartition limit, $\gamma = 1+2/f$, where $f$ is the number of atomic or molecular degrees of freedom. This together yields the approximate expression for the adiabatic equation
\begin{equation}
    T(p) = T_0 \left(\frac{p}{p_0} \right)^{R/c_p} 
    %= \left(\frac{p}{p_0} \right)^{{\gamma-1}/\gamma}
    \approx \left(\frac{p}{p_0} \right)^{2/(f+2)}
\end{equation}
where $p$ is pressure, $T$ temperature, and $p_0$ and $T_0$ are reference values. 
We have $f = 3$ for monoatomic He and $f = 5$ for diatomic \ce{H2} and \ce{N2}. Hence the exponent is approximately 2/5 for He and 2/7 for \ce{H2} and \ce{N2}. Given similar emission temperatures, the result is a higher surface temperature for \ce{He} vs. \ce{N2} composition \citep{pierrehumbert2010principles}; around 17~K higher in the G dwarf case and 12~K higher in the M dwarf case (Figure~\ref{fig:1}).

\begin{figure*}[ht!]
\centering
\includegraphics[scale = 0.3]{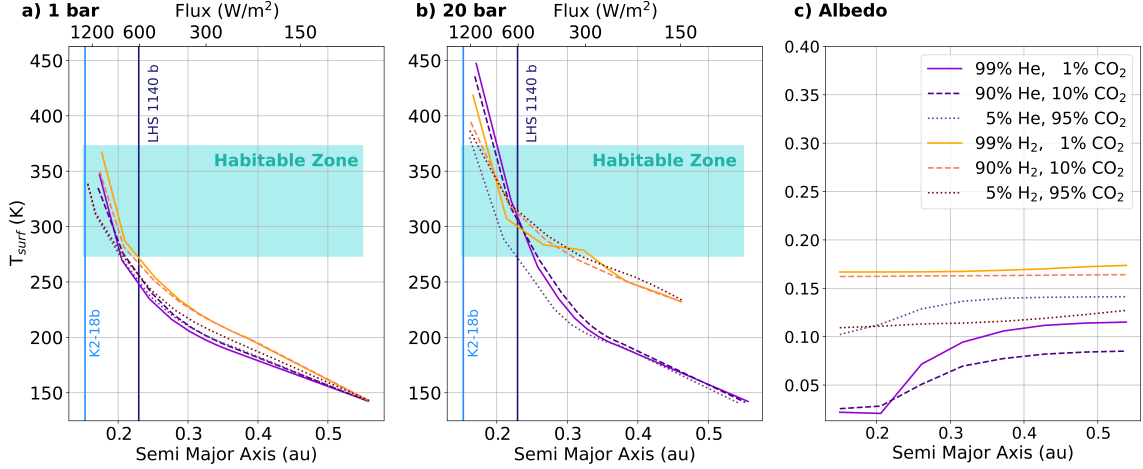}
\caption{Surface temperature (assuming an M dwarf spectrum and $L = 0.0234 \: L_{Sun}$, resembling K2-18) as a function of semi-major axis (bottom x-axis) and incident flux at the planet (top x-axis), respectively. The blue-shaded area represents the region where liquid surface water is possible. The blue vertical lines indicate the received fluxes of K2-18b and LHS 1140 b, as reported by \citet{benneke2019water} and \citet{cadieux2024new}, respectively. \textbf{a)} Varying molar concentrations of \ce{CO2} (1\%, 50\%, 95\%) with He and \ce{H2}, respectively, as background gas with $p_s$ = 1 bar. \textbf{b)} Varying molar concentrations of \ce{CO2} (1\%, 50\%, 95\%) with He and \ce{H2}, respectively, as background gas with $p_s$ = 20 bar. \textbf{c)} Albedo as a function of semi-major axis for varying molar concentrations of \ce{CO2} (1\%, 50\%, 95\%) with He and \ce{H2}, respectively, as background gas with $p_s$ = 20 bar.}
\label{fig:2}
\end{figure*}

Results for varying stellar flux as a function of semi-major axis are shown in Figure \ref{fig:2}, for the M dwarf case only, given 1-bar and 20-bar background pressures. As can be seen, temperatures permitting liquid water are possible at orbital distances shorter than 0.2~au in the 1~bar case, and in the range of 0.2~au to 0.25~au in the 20~bar case. Time-evolving temperature profiles fail to determine the radiative-convective equilibrium for the hottest cases we studied due to their high atmospheric opacities, causing the surface temperature to decouple from the upper atmosphere. As a result, the temperature rises indefinitely and they become numerically unstable. So for $d<0.2$~au, we calculated surface temperature using fixed moist adiabatic temperature profiles with stratospheric temperature of 150~K.

In the 1 bar scenario, the 95\% cases of both \ce{H2} and He converge, whereas in the 20 bar case the warming effect of the \ce{H2} atmospheres due to CIA dominates.
In the 20 bar scenario, the He-dominated atmosphere is much warmer than the 5$\%$ He, 95$\%$ \ce{CO2} atmosphere at small orbital distances (more than 50 K difference in $T_{surf}$ at 0.2 au). The same is true for the \ce{H2} atmospheres at the inner edge, but to a smaller extent. This may be due to the uniquely large scale height and water-vapor column of a \ce{H2} atmosphere, which reduces OLR and increases surface temperature \citep{koll2019hot}.  \newline
Moreover, the He-dominated atmospheres lead to higher $T_{surf}$ at the inner edge compared to \ce{H2}-dominated atmospheres. This is due to the extremely low albedo of the He-dominated case compared to the \ce{H2}-dominated atmospheres at the inner edge (0.021 vs. 0.17 at 0.2 au, see Figure \ref{fig:2}). The \ce{H2}-dominated atmospheres have the highest albedo due to stronger Rayleigh scattering per unit column compared to \ce{CO2} and \ce{He}. 
The increase in albedo at intermediate orbital distances in the 99\% and 90\% \ce{He} cases is a 
consequence of variation of the water concentration. Absorption by water vapor dominates over scattering by the background gases at low orbital distances due to He's weak Rayleigh scattering, but as surface temperature decreases, this effect becomes weaker allowing Rayleigh scattering due to He to dominate. 
Below 200~K, the absorption by water vapor is negligible. This was validated by modeling a 100\% \ce{He} atmosphere with condensable \ce{H2O}, which resulted in an almost transparent atmosphere below 200~K. This caused the planetary albedo to approach the surface albedo, and the surface temperature to be close to the value expected on an airless planet.
The He-dominated atmospheres also enter a runaway greenhouse regime sooner at close orbital distances, due to their lower albedos. This result should be seen as an upper limit on the He-dominated runaway greenhouse transition, because our clear-sky calculations neglect the reflective effects of clouds and aerosols. \newline
If an atmosphere contains species that have a higher molecular weight than the main atmospheric component, they may enter the Guillot regime \citep{guillot1995condensation, seeley2025resolved}. If the Guillot threshold is crossed, warmer air packages no longer sustain buoyancy due to their high relative weight relative to the background gas. As a result, convection is suppressed and heat transport is dominated by radiation, leading to the formation of a radiative, superadiabatic layer \citep{habib20253d}. The threshold can be calculated according to eq. (13) in \citet{seeley2025resolved}.
For the \ce{He}-dominated atmospheres (both 1~bar and 20~bar), the Guillot regime may affect our results for received stelalr flux values larger than around 750~\si{W.m^{-2}}, and for the 1~bar \ce{H2}-dominated atmospheres for flux values larger than around 460~\si{W.m^{-2}}. At 20~bar, the \ce{H2}-dominated atmospheres may exhibit suppressed convection at all orbital distances. The PCM Line-By-Line code does not explicitly handle convection in the Guillot regime, so this question is left to future work.

In summary, He-dominated atmospheres have steeper tropospheric adiabats compared to atmospheres with di- or triatomic background gases. In the clear-sky case, they exhibit a narrower habitable zone compared to \ce{H2}-dominated atmospheres,  with the inner edge  shifting outward and the outer edge inward. At the inner edge, this shift is due to the  lower albedo of \ce{He} atmospheres. At the outer edge, the more rapid freezing occurs due to the lack of strong CIA absorption in key infrared spectral regions compared to \ce{H2}.

\begin{figure*}[ht!]
\centering
\includegraphics[scale=0.55]{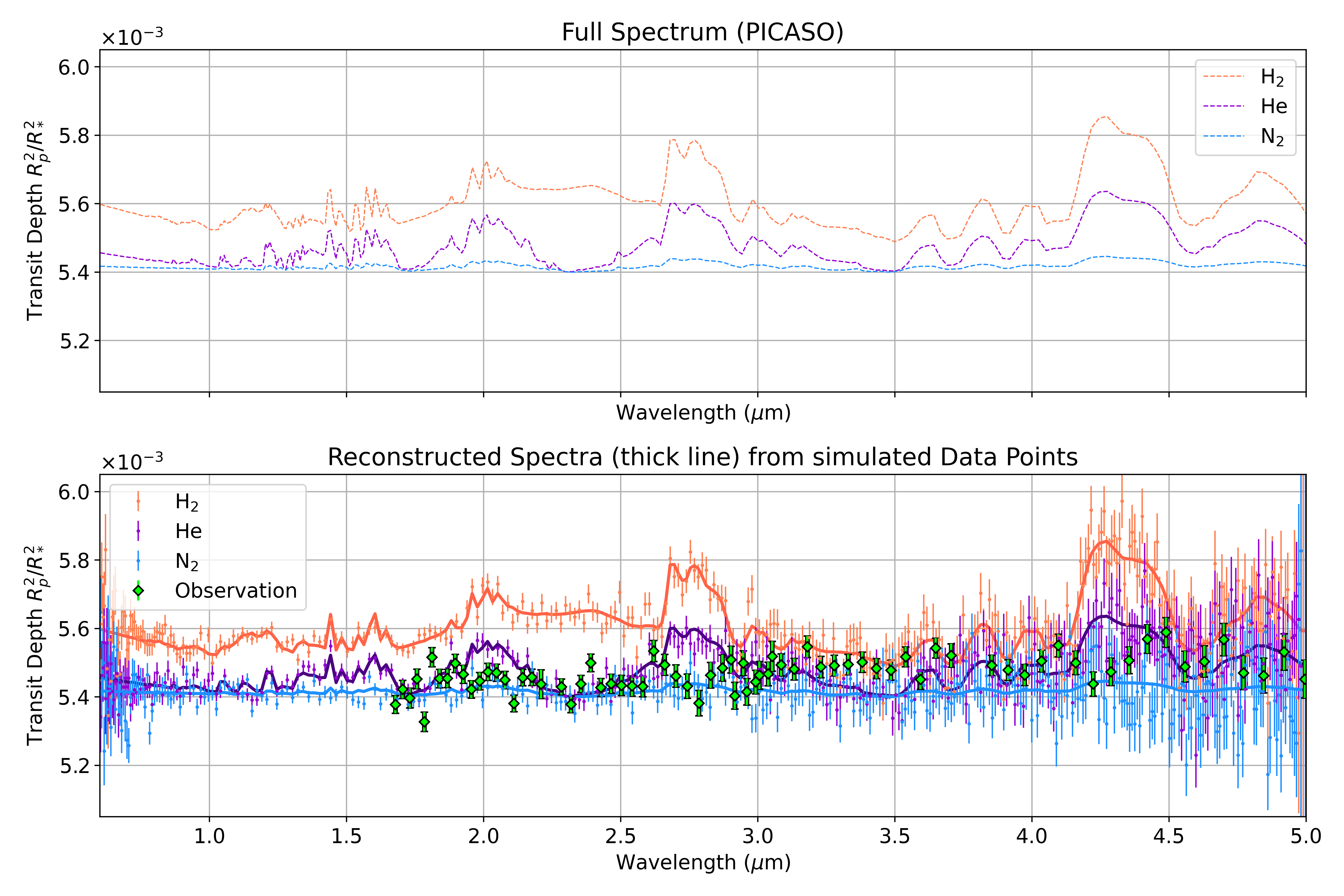}
\caption{\textbf{Top Panel:} Transmission spectra generated using PICASO  \citep{batalha2019exoplanet} for atmospheres dominated by \ce{N2}, \ce{H2}, He with 1$\%$ CO$_2$ levels and condensable \ce{H2O}.
The planetary and stellar properties are those of exoplanet LHS 1140 b as described by \citet{cadieux2024new} and \citet{cherubim2025oxidation}.
\textbf{Bottom Panel:} Simulated JWST data using PandExo (\citet{batalha2017pandexo}) for the same atmospheres. The points represent the simulated data points with the corresponding error bars. The thicker lines represent the reconstructed transmission spectra based on the simulated data in each case. The instrument chosen for this simulation was NIRSpec PRISM. Observational data of LHS 1140 b reported by \citet{damiano2024lhs} is shown in green.}
\label{fig:4}
\end{figure*}
\vspace{-3mm}

Despite the narrow range of received stellar fluxes and surface pressures over which He-dominated atmospheres permit surface liquid water, they are favorable targets for transit spectroscopy. This can be seen in Figure \ref{fig:4}, which shows a transmission spectrum generated using \verb|PICASO| for \ce{H2}, \ce{He} and \ce{N2} dominated atmospheres, alongside a JWST detectability analysis employing \verb|PandExo|. Additionally, spectroscopy results for LHS 1140 b reported by \citet{damiano2024lhs} are overplotted onto the modeled spectra. Both \citet{damiano2024lhs} and \citet{cadieux2024new} argued against the possibility of hazes in LHS 1140 b's atmosphere based on \ce{H2}-atmosphere modeling by \citet{hu2021photochemistry}. However, more modeling with He as the background gas is necessary to fully exclude the possibility of e.g. sulfur-based hazes as a potential cause of the discrepancy between observational data of LHS 1140 b's atmosphere and the modeled spectra. Due to the much lower scale height in the \ce{N2}-dominated case [14.6~km (\ce{N2}) versus 106.5~km (\ce{H2}) and 70.5~km (He)], its spectrum is almost flat. In contrast, the spectra of the \ce{H2} and He atmospheres have significantly deeper features and the 2.0$\mu$m, 2.7$\mu$m and 4.3$\mu$m absorption lines of CO$_2$ are clearly recognizable in their reconstructed spectra. This indicates He-dominated atmospheres will be significantly easier to characterize with infrared spectroscopy than atmospheres dominated by \ce{N2}, or other heavy species such as \ce{CO2}.
\newline

\section{Discussion} \label{sec:discussion}

We have focused on extremely low mass fraction He-dominated atmospheres here for modeling convenience and to elucidate the key physical features. Such atmospheres are possible as the result of fractionation during mass loss of a primordial \ce{H2}-\ce{He} envelope \citep{hu2015helium,malsky2023helium,cherubim2024strong,lammer2025earth}. While they are likely to be transient, due to continued He loss to space over time, they can remain stable on Gyr timescales \citep{cherubim2025oxidation}. The erosion timescales can be estimated using the upper limit on mass flux

\begin{equation}
    \phi_{XUV} = \frac{\epsilon F_{XUV}}{4 V_{pot}},
\end{equation}

with $\epsilon \approx 0.15$ and $V_{pot}$ the gravitational potential \citep{cherubim2024strong}. For the test case of LHS 1140 b, we adopt the semi-empirical estimate by XMM-Newton for the stellar $F_{XUV}$ of LHS 1140, $F_{XUV} = 0.033~Wm^{-2} - 0.17~Wm^{-2}$ (Cherubim et al. 2026, in preparation). We obtain erosion timescales of 5.4~Myr to 27.6~Myr for the 1~bar case, and 107.2~Myr to 552.3~Myr for the 20~bar case. These values likely constitute lower limits on the atmospheric erosion timescales, as they neglect interactions with the planetary interior, and radiative cooling by metal-rich species. 

\begin{figure*}[ht!]
\centering
\includegraphics[scale=0.35]{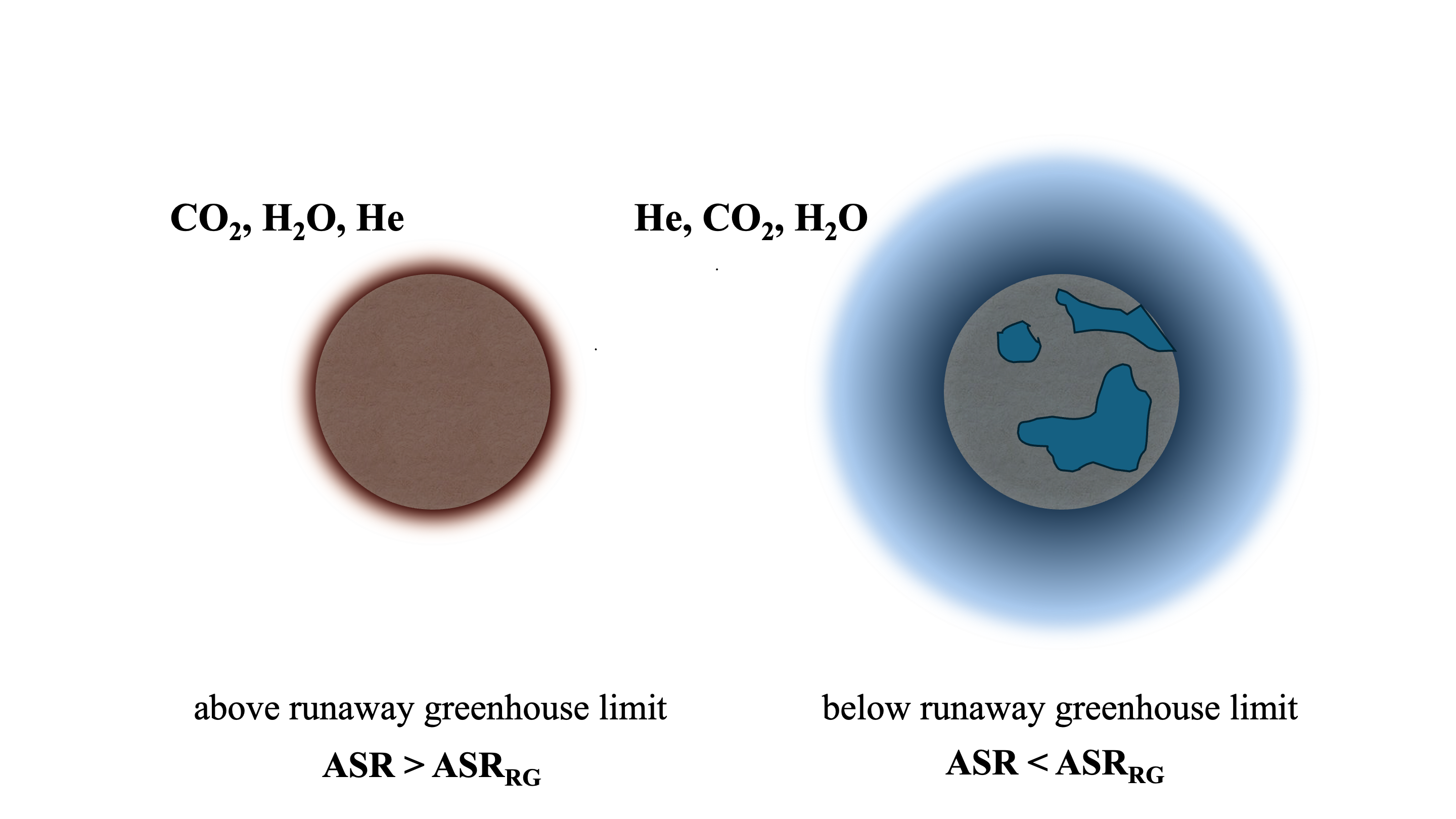}
\caption{Cartoon illustrating the effect of the runaway greenhouse transition on atmospheric composition for two rocky planets with identical He, \ce{H2O} and \ce{CO2} inventories. With a thin He atmosphere, planets below the runaway greenhouse limit (right) will have He-rich upper atmospheres and may support surface liquid water. For planets inside the inner edge of the habitable zone (left), \ce{H2O} from liquid water oceans and \ce{CO2} from carbonates are transported into the atmosphere, dramatically decreasing the scale height and making the transit spectrum of the planet flatter.}
\label{fig:5}
\end{figure*}

For He-rich planets with Earth-like or larger \ce{H2O} and \ce{CO2} inventories, the runaway greenhouse transition would likely lead to a substantial increase in mean molecular mass in the atmosphere, due to conversion of liquid water oceans and surface carbonate deposits to \ce{H2O} vapor and gaseous \ce{CO2}, respectively. This could conceivably lead to population-level differences in the transit spectra of rocky exoplanets in the He-enrichment part of the radius valley (Figure~\ref{fig:5}). It would be interesting to explore this possibility further in future work.

It will also be interesting to extend our analysis to higher pressure cases in future work, as these are expected to be more common. Above 20~bar, the Guillot effect \citep{seeley2025resolved} as well as the effects of supercritical interiors \citep{pierrehumbert2023runaway} will play an important role and be integral to an accurate analysis. Future studies should also investigate the effects of clouds and aerosols on both albedo and infrared radiation in He-dominated atmospheres.

\section{Conclusions} \label{sec:conclusion}

We presented radiative-convective and transmission spectrum modeling results for rocky exoplanets with He-dominated atmospheres using the open-source codes PCM-LBL, \verb|PICASO| and \verb|PandExo|. Our key conclusions are:

\begin{enumerate}
    \item He atmospheres exhibit a steeper temperature gradient in the troposphere compared to atmospheres dominated by diatomic gases, which leads to a stronger greenhouse effect when \ce{CO2} and \ce{H2O} are also present.
    \item He exhibits lower Rayleigh scattering than diatomic gas species, but line broadening differences among \ce{He}, \ce{H2} and \ce{N2} are small.
    \item Liquid surface water is possible in \ce{He}-dominated atmospheres, but for a narrower range of orbital distances than for \ce{H2} atmospheres.
    \item Cloud-free He-dominated atmospheres containing \ce{CO2}, \ce{H2O} or similar infrared absorbers on rocky M dwarf planets like LHS 1140 b are detectable by JWST. 
\end{enumerate}

\begin{acknowledgments}
This work was carried out in affiliation with ETH Zurich and Harvard University, and V.K. thanks the Department of Earth and Planetary Sciences at Harvard University for hospitality and support during the course of this work. R.W. and J. C. acknowledge support from Leverhulme Center for Life in the Universe grant G119167, LBAG/312 and NSF grant AGS-2210757. We thank Iouli Gordon for discussion of the radiative properties of He-dominated atmospheres and opacity data on He-\ce{CO2} CIA. We also thank Jake Seeley for sharing his expertise on the Guillot effect.
\end{acknowledgments}

\software{ PCM--LBL \citep{wordsworth2021coupled}, PICASO \citep{batalha2019exoplanet}, PandExo \citep{batalha2017pandexo}}

The data and code used to generate the figures in this work are available at DOI \verb|10.5281/zenodo.20788580|.

\bibliographystyle{apalike}
\bibliography{sample7}

\end{document}